\documentclass[twocolumn,showpacs,preprintnumbers,superscriptaddress,prb,floatfix,aps,10pt]{revtex4-2}

\usepackage{hyperref}
\hypersetup{
    bookmarks=true,         % show bookmarks bar?
    unicode=true,           % non-Latin characters in Acrobat’s bookmarks
    pdftoolbar=false,       % show Acrobat’s toolbar?
    pdfmenubar=true,        % show Acrobat’s menu?
    pdffitwindow=true,      % page fit to window when opened
    pdftitle={My title},    % title
    pdfauthor={Author},     % author
    pdfsubject={Subject},   % subject of the document
    pdfnewwindow=true,      % links in new window
    pdfkeywords={keywords}, % list of keywords
    colorlinks=true,        % false: boxed links; true: colored links
    linkcolor=blue,         % color of internal links
    citecolor=blue,         % color of links to bibliography
    filecolor=black,        % color of file links
    urlcolor=blue           % color of external links
  }

\usepackage{amsmath,amssymb}
\usepackage{graphicx,textcomp}
\usepackage{subfigure}
\usepackage{color}
\usepackage{todonotes}
\usepackage[titletoc,toc,title]{appendix}
\usepackage{xcolor}
\usepackage{todonotes}
\usepackage[utf8]{inputenc}
\usepackage{lipsum}

\definecolor{bjorn}{RGB}{240,226,182}
\definecolor{davide}{RGB}{211,22,112}

\newcommand{\figref}[1]{Fig.~\ref{#1}}
\newcommand{\secref}[1]{Sec.~\ref{#1}}
\newcommand{\eqnref}[1]{Eq.~\ref{#1}}

\newcommand{\angstrom}{\text{\normalfont\AA}}

\begin{document}

\title{Phase stability of Fe from first-principles: atomistic spin dynamics coupled with ab initio molecular dynamics simulations and thermodynamic integration}
\author{Davide Gambino}
\email{davide.gambino@liu.se}
\affiliation{Department of Physics, Chemistry, and Biology (IFM), Link\"{o}ping University, SE-581 83, Link\"oping, Sweden.}

\author{Johan Klarbring}
\altaffiliation[Present address: ]{Department of Materials, Imperial College London, Exhibition Road, London SW7 2AZ, UK}
\affiliation{Department of Physics, Chemistry, and Biology (IFM), Link\"{o}ping University, SE-581 83, Link\"oping, Sweden.}

%\affiliation{Department of Materials, Imperial College London, Exhibition Road, London SW7 2AZ, UK}

\author{Bj\"{o}rn Alling}
\affiliation{Department of Physics, Chemistry, and Biology (IFM), Link\"{o}ping University, SE-581 83, Link\"oping, Sweden.}

\date{\today}

\begin{abstract}

The calculation of free energies from first principles in materials is a formidable task which enables the prediction of phase stability with high accuracy; these calculations are complicated in magnetic materials by the interplay of electronic, magnetic, and vibrational degrees of freedom.
In this work, we show the feasibility and accuracy of the calculation of phase stability in magnetic systems with $ab$ $initio$ methods and thermodynamic integration by sampling the magnetic and vibrational phase space with coupled atomistic spin dynamics-$ab$ $initio$ molecular dynamics (ASD-AIMD) simulations [Stockem $et \; al.$, PRL \textbf{121}, 125902 (2018)], where energies and interatomic forces are calculated with density functional theory (DFT).
We employ the method to calculate the phase stability of Fe at ambient pressure from 800 K up to 1800 K.
The Gibbs free energy difference between fcc and bcc Fe at zero pressure as a function of temperature is calculated carrying out thermodynamic integration over temperature on the average energies at the DFT level obtained by ASD-AIMD. 
The thermodynamic integration over temperature needs a reference free energy difference, which we directly calculate in the paramagnetic state at temperatures much higher than the magnetic transition temperatures with thermodynamic integration over stress-strain variables using disordered local moment (DLM)-$ab$ $initio$ molecular dynamics simulations.
For bcc Fe, we employ two different sets of exchange interactions as a function of distance in the ASD-AIMD procedure and we show the importance of the magnetic ordering temperature on the $\alpha$ to $\gamma$ structural transition temperature, whereas the $\gamma$ to $\delta$ transition is well reproduced independently of the exchange interactions.
For the best set of exchange interactions, the Gibbs free energy difference between the two structures is within 5 meV/atom from the CALPHAD estimate, and both transition temperatures are reproduced within 150 K.
The present work paves the way to free energy calculations in magnetic materials from first principles with accuracy in the order of 1 meV/atom.

\end{abstract}
\maketitle

\section{Introduction}

The prediction of temperature-dependent phase stability in materials from first principles is one of the great challenges of the electronic structure community \cite{PhaseStabilityFirstPrinciples}.
Phase stability is governed by the Gibbs free energy, whose minimum dictates the equilibrium phase of a material at constant pressure and temperature.
The calculation of the Gibbs free energy from first principles can be carried out with thermodynamic integration (TI) methods, which relate free energies with other thermodynamic quantities accessible from atomistic simulations \cite{Frenkel}.
Examples of such calculations with first principles methods are present in the literature \cite{AlfeTI,Al-UPTILD,SSTIKlarbring,SiMeltRPA,ExcComparisonTILD}, showing that this field is now mature.
TI requires exploration of the phase space of a system, which can be carried out with $ab$ $initio$ molecular dynamics (AIMD) simulations, where interatomic forces and energies are calculated with density functional theory (DFT).
AIMD simulations are a fundamental tool in the determination of free energies, although imperiled by several sources of errors such as the underlying exchange and correlation functional \cite{ExcComparisonTILD}, finite size of the simulation cell, time of the simulations, and employed thermostats, to name a few.
Nonetheless, the combination of TI and AIMD has been proven to be able to model phase stability in real systems \cite{Al-UPTILD,SSTIKlarbring,SiMeltRPA,ExcComparisonTILD} and it is therefore the most accurate method currently at hand.

However, the previously cited studies are carried out in nonmagnetic systems, with the exception of Ni in Ref. \cite{ExcComparisonTILD}.
In fact, magnetism in materials adds a further level of complexity in computer simulations, since the phase space of the material includes also the magnetic degree of freedom (DOF).
It is well known by now that the change in magnetic state can lead to changes in the interatomic force constants in some materials like bcc Fe \cite{SSA_bccFe,phononsFeDMFT}, and finite temperature calculations need to take into account short range order effects \cite{phononsSSAwSRO} in order to have quantitative predictive power, underlining the need of treating consistently the magnetic and vibrational DOF in AIMD simulations.
Spin-lattice dynamics \cite{SLD_Ma_Dudarev,SLD_Uppsala} is a method that evolves the magnetic and vibrational DOF in a consistent way, although it is based on interatomic potentials, therefore inheriting all the issues of accuracy and transferability related with these potentials.
The interatomic potentials can be fitted to DFT data, which improves strongly the accuracy of these methods.
A step forward toward a full $ab$ $initio$ treatment of the coupled magneto-lattice dynamics was taken with the development of combined atomistic spin dynamics-$ab$ $initio$ molecular dynamics (ASD-AIMD) simulations \cite{ASD-AIMD}.
In this method, the forces between atoms are calculated with DFT, and the evolution of the magnetic moments is simulated with spin dynamics simulations based on Landau-Lifshitz-Gilbert equations.
% The ASD and the AIMD parts communicate in this method: the ASD part dictates the directions of the moments at each AIMD step, therefore influencing the interatomic forces; the AIMD part influences instead the pair exchange interactions, which on first approximation depend on the distance between the atoms, and therefore the strength of interaction between the magnetic moments in the ASD part.
ASD-AIMD simulations in CrN, an antiferromagnetic semiconductor, showed an anomalous scattering of phonons  due to coupling between the magnetic and lattice DOF just above the N\'eel temperature, which could not be explained with simulations in an ideal paramagnetic (PM) state carried out with disordered local moment (DLM) \cite{Hubbard_I,*Hubbard_II,*Hubbard_III,Hasegawa_I,*Hasegawa_II,DLM_Gyorffy,DLM_MSM_Alling} molecular dynamics (DLM-AIMD) method \cite{DLMMD_Steneteg}.
This success of ASD-AIMD simulations in the description of a magneto-lattice coupling effect suggests that, in combination with TI methods,  the calculation of temperature-dependent phase stability also in magnetic materials from first principles is now an achievable goal.

An ideal system to test such a combination of techniques is Fe, the archetype of magnetic metals.
Indeed, Fe shows a sequence of phase transitions occurring at ambient pressure, making it a suitable candidate to benchmark free energy methods.
At low temperatures, Fe is in a bcc ferromagnetic (FM) phase, the $\alpha$ phase; at 1043 K, the system undergoes a magnetic transition to PM bcc Fe, closely followed by a structural transition to fcc ($\gamma$ phase) at 1185 K; finally, at 1667 K, the system turns back to the bcc structure, known as the $\delta$ phase, before melting at 1811 K.
The experimentally estimated Gibbs free energy difference between bcc and fcc Fe in the stability range of the $\gamma$ phase is estimated to be in the order of 1 meV/atom \cite{FritzThesis}, requiring great accuracy for any technique employed to calculate this quantity if the $\alpha-\gamma-\delta$ transition is to be captured.
The calculation of free energies in Fe was reported in Ref. \cite{Ma_Dudarev_Fe_transition}, where spin-lattice dynamics was employed for the sampling of the phase space.
The results from that study were in very good agreement with the experimental findings, although a tweaking of the exchange interactions was needed to achieve this impressive result.
Recently, the phase stability in Fe was also calculated with a tight binding Hamiltonian and thermodynamic integration\cite{TIBochum}.
Another study \cite{Fe_Bjorn} showed using DLM-AIMD simulations that bcc and fcc Fe at the $\gamma \rightarrow \delta$ transition have very similar electronic and magnetic properties, suggesting that the transition is driven by the larger vibrational entropy of the bcc structure compared to the fcc structure.
However, this study neglected any short-range order effect at finite temperatures, which is governed by the exchange interactions $J_{ij}$.

Much research has been done on the exchange interactions in bcc Fe.
Several studies have calculated the exchange interactions in this system with different methods\cite{MFT_II,FeJijMFT,SGPM,YinJijVib,RubanJijVib}.
For an accurate description of the transition temperature, the reference magnetic phase is of great importance \cite{SGPM}: since at the Curie temperature the system is more similar to a PM state than a perfect FM state, the former state should be used as background in the calculation of the $J_{ij}$.
Later on, a strong effect of lattice vibrations on the value of the exchange interactions \cite{YinJijVib,RubanJijVib} was found, which is responsible for a reduction of the theoretical transition temperature to values lower than the experimentally determined value.
Less focus was given to the fcc phase of Fe, which shows a more complex magnetism than bcc Fe: it has been proven that the ground state of fcc Fe is a spin-spiral arrangement \cite{fccFeGS}, but for larger volumes the system becomes FM \cite{fccFeJijtransition}.
Up to our knowledge, no study of the effect of vibrations on the exchange interactions has been carried out on fcc Fe.
In addition to the difficulties in the modeling of exchange interactions, a further level of complexity is introduced by the longitudinal degrees of freedom in metallic magnets, also known as longitudinal spin fluctuations (LSF).
Several methods have been developed in the years to account for this effect on a semiclassical level \cite{LSFMurata,LSFKubler,LSFRosengaard,LSFRuban,classicalLSFWysocki,LSFRubanEarthCore,LSFKorzhavyiFe,LSFVitos,LSFKhmelevskyi,LSFGambino}; an efficient method to include this effect in DFT calculations was introduced in Ref. \cite{LSFRubanEarthCore}.

In this work, we attempt to include the coupling of all these effects within the same first principles framework in the calculation of phase stability in Fe at zero pressure.
We carry out the calculation of the fcc-bcc Gibbs free energy difference in Fe as a function of temperature with the use of ASD-AIMD and DLM-AIMD simulations and thermodynamic integration.
This task is achieved by the definition of a thermodynamic path that starts from the direct calculation of the free energy difference between the two structures at high temperature in the DLM state with the stress-strain thermodynamic integration method (SSTI) \cite{Wallace,SSTI_example_III,SSTIKlarbring} along a deformation path of the Bain type.
This free energy difference is then used as reference to obtain the full temperature dependence with thermodynamic integration over temperature (TTI) and, including free energy contributions from thermal expansion, the full Gibbs free energy difference between fcc and bcc Fe as a function of temperature at zero pressure is retrieved including electronic, magnetic and vibrational parts and their coupling.

Our work starts by introducing the ASD-AIMD and DLM-AIMD methods (Sec. \ref{sec:ASDAIMD-DLMMD}), together with the computational details employed for these calculations. 
The exchange interactions employed in this work are shown in Sec. \ref{sec:Jij}.
In Sec. \ref{sec:TIpath}, the thermodynamic path and the TTI are presented, followed by the calculation of equilibrium volumes in Sec. \ref{sec:VcorrF} and the SSTI in Sec \ref{sec:SSTI}.
The calculated constant-volume free energy difference and the Gibbs free energy difference at zero pressure are presented in Sec. \ref{sec:Results} and compared with literature.
In Sec. \ref{sec:Conclusions}, the conclusions of this study are drawn.

%\section{Methods} \label{sec:methods}

\section{ASD-AIMD and DLM-AIMD simulations: methods and computational details} \label{sec:ASDAIMD-DLMMD}

ASD-AIMD simulations, introduced first in Ref. \cite{ASD-AIMD},
consist in an ASD and an AIMD simulation run in parallel while intercommunicating with each other.
At each AIMD timestep, interatomic forces are calculated with DFT employing the direction of the magnetic moments from ASD simulations, so that the magnetic DOF influence the evolution of the atomic positions.
At the same time, the different atomic positions at each step determine different exchange interactions between neighboring moments, which therefore affect the evolution of the ASD part of the simulation.
The parametrized, distance-dependent pair exchange interactions employed in this work are shown in Sec. \ref{sec:Jij}.
Since in the ASD part the moments are 3D vectors, we employ noncollinear DFT calculations at each AIMD step.
In order to keep the moments in the direction dictated by the ASD simulation, we use the constraining method developed by Ma and Dudarev \cite{Ma_Dudarev_constraint}.

In addition, we use an LSF term of the type $S^{\textrm{LSF}}=-k_B \log (m)$ in our VASP calculations,
with $m$ being the size of the magnetic moment. 
This expression can be derived from a semiclassical thermodynamic model of LSF with monodimensional phase space measure \cite{classicalLSFWysocki,LSFKhmelevskyi,LSFGambino}.
The LSF entropic term is directly introduced in the Kohn-Sham single-particle potentials, as first done in Ref. \cite{LSFRubanEarthCore}.
One uncertainty related to this term consist in the nature of the systems under investigation: the monodimensional phase space measure assumed in this work is suited for localized moment systems, as bcc Fe, whereas a different expression can be derived for a more itinerant system \cite{LSFRubanEarthCore}.
Fcc Fe is often considered a more itinerant system than bcc Fe, nonetheless for the volumes considered here the the Fe atoms in the fcc structure are in the high-spin state, more similar to bcc Fe.
We therefore employ the same LSF term for both bcc and fcc Fe, as well as for the intermediate structures along the deformation path.
The magnetic entropy is not included in the calculation of the stresses, therefore we cannot include it as an adiabatic term, and we thus include only its indirect effect of increasing the size of the magnetic moments, rather than as a separate free energy term. 
By stabilizing slightly larger moments, computational gains are observed as the convergence of DFT ionic steps is made easier.

DLM-AIMD simulations were first developed in Ref. \cite{DLMMD_Steneteg} in a collinear framework, where the AIMD run was performed by changing the magnetic configuration every few timesteps.
The magnetic configurations consisted of a random distribution of up and down moments.
In this work, we perform DLM-AIMD simulations in a different way, to ensure compatibility with the ASD-AIMD simulations just described.
We employ here the same computational tools of the ASD-AIMD simulations, but in the ASD-part of the simulation we set exchange interactions to 0 and magnetic temperature to $10^6$ K.
The LSF temperature is always set to the temperature of the AIMD part.
In this way, we ensure that the spin-spin correlation functions are close to zero at each step, and the spin autocorrelation function goes to zero in one single timestep, modeling an ideal, adiabatically fast PM state.

To ensure computational efficiency, we carry out the simulations for $8-12$ ps with an AIMD timestep of 1 fs and low accuracy parameters, namely $\Gamma$-point sampling of the Brillouin zone and cutoff for the plane-wave expansion of 250 eV.
Snapshots from each simulation are then upsampled with higher convergence parameters, $2\times 2\times 2$ k-points mesh and 500 eV cutoff; the upsampled snapshots are separated by approximately 100-200 fs from each other, to ensure statistical independence.
Throughout the paper, the reported energies and stresses are obtained by summing the average quantity from the low accuracy calculation with the average difference in the quantity between the high and low accuracy calculations, as suggested in Ref. \cite{SiMeltRPA}.
We checked the validity of the upsampling scheme by comparing the average energy and stresses obtained with this method with the results of a full simulation with the high convergence parameters for the bcc structure at 1000 K and a single volume with ASD-AIMD, and for bcc, fcc and one intermediate structures at 1800 K with DLM-AIMD 

AIMD simulations are carried out in the canonical ensemble, employing the Langevin thermostat to control temperature.
The timestep for the AIMD part of the simulations, as mentioned, is set to 1 fs, whereas the ASD simulations between each AIMD step are performed for the same duration of 1 fs but with a smaller timestep, $10^{-2}$ fs.
Each run is started with a thermalization period of 500-1000 timesteps, which are discarded.
Simulations are performed with body-centered tetragonal (bct) structures, which for particular values of the $\textbf{c}$ lattice vector correspond to bcc and fcc structure.
In particular, the matrices of the lattice vectors for the bcc and fcc structures are:
\begin{equation}\label{eq:bccfccmatrices}
    \begin{split}
    &\textbf{h}^{\textrm{bcc}}=
    \begin{pmatrix}
        a^{\textrm{bcc}} & 0   & a^{\textrm{bcc}}/2\\
        0   & a^{\textrm{bcc}} & a^{\textrm{bcc}}/2\\
        0   & 0   & a^{\textrm{bcc}}/2
    \end{pmatrix}\\[5pt]
    &\textbf{h}^{\textrm{fcc}}=
    \begin{pmatrix}
        a^{\textrm{fcc}} & 0   & a^{\textrm{fcc}}/2\\
        0   & a^{\textrm{fcc}} & a^{\textrm{fcc}}/2\\
        0   & 0   & a^{\textrm{fcc}}/\sqrt{2}
    \end{pmatrix}.
    \end{split}
\end{equation}
The columns of these matrices correspond to the lattice vectors of the cell; the lattice parameter $a^{\textrm{bcc}}$ corresponds to the lattice parameter in the conventional bcc cell, whereas the lattice parameter $a^{\textrm{fcc}}$ is reduced of a factor of $1/\sqrt{2}$ compared to the lattice parameter of the conventional fcc cell.
These definitions of the cells are particularly useful because they can be transformed one into the other along the Bain path \cite{AlfeTI}.
We use supercells made of $5\times 5\times 5$ repetitions of the bct cells, for a total of 125 atoms.

All DFT calculations were carried out with the Vienna $ab$ $initio$ simulation package (VASP) \cite{VASP_I,*VASP_II,*VASP_III} with the projector augmented wave (PAW) method \cite{PAW_Blochl,PAW_vasp} and plane-waves as basis set.
Convergence for the electronic optimization was set to $10^{-5}$ eV.
The sampling of the Brillouin zone was performed with the Monkhorst-Pack scheme \cite{MPscheme}.
Details on the k-points meshes and cutoff for expansion over the plane-wave basis where already given in the relevant sections.
We include the electronic entropic contribution through the Mermin functional \cite{MerminFunctional} using the Fermi-Dirac electronic smearing and electronic temperature corresponding to the simulated temperature. 
ASD simulations were performed with the UppASD code \cite{UppASD_I,UppASD_II} using a phenomenological damping factor of 0.05.

Recently, the calculation of stresses in a constrained DFT framework has been carried out for bcc Fe with the two moments in the conventional bcc cell at arbitrary angle with respect to each other \cite{cDFTstresses}, suggesting that an additional term appears in the stresses due to the constraints.
We have replicated the calculation of the pressure upon magnetic moment rotations performed in Ref. \cite{cDFTstresses} and we only observe differences within 10 kbar, comparable with the differences in magnetic moment sizes obtained.
The error introduced by the suggested missing term of the stresses due to constraints is small and probably in the order of the numerical accuracy of the present calculations.

\section{Distance-dependent exchange interactions} \label{sec:Jij}

\begin{figure}%
\centering
\subfigure{\label{fig:bccJijR} \includegraphics[width=0.4\textwidth]{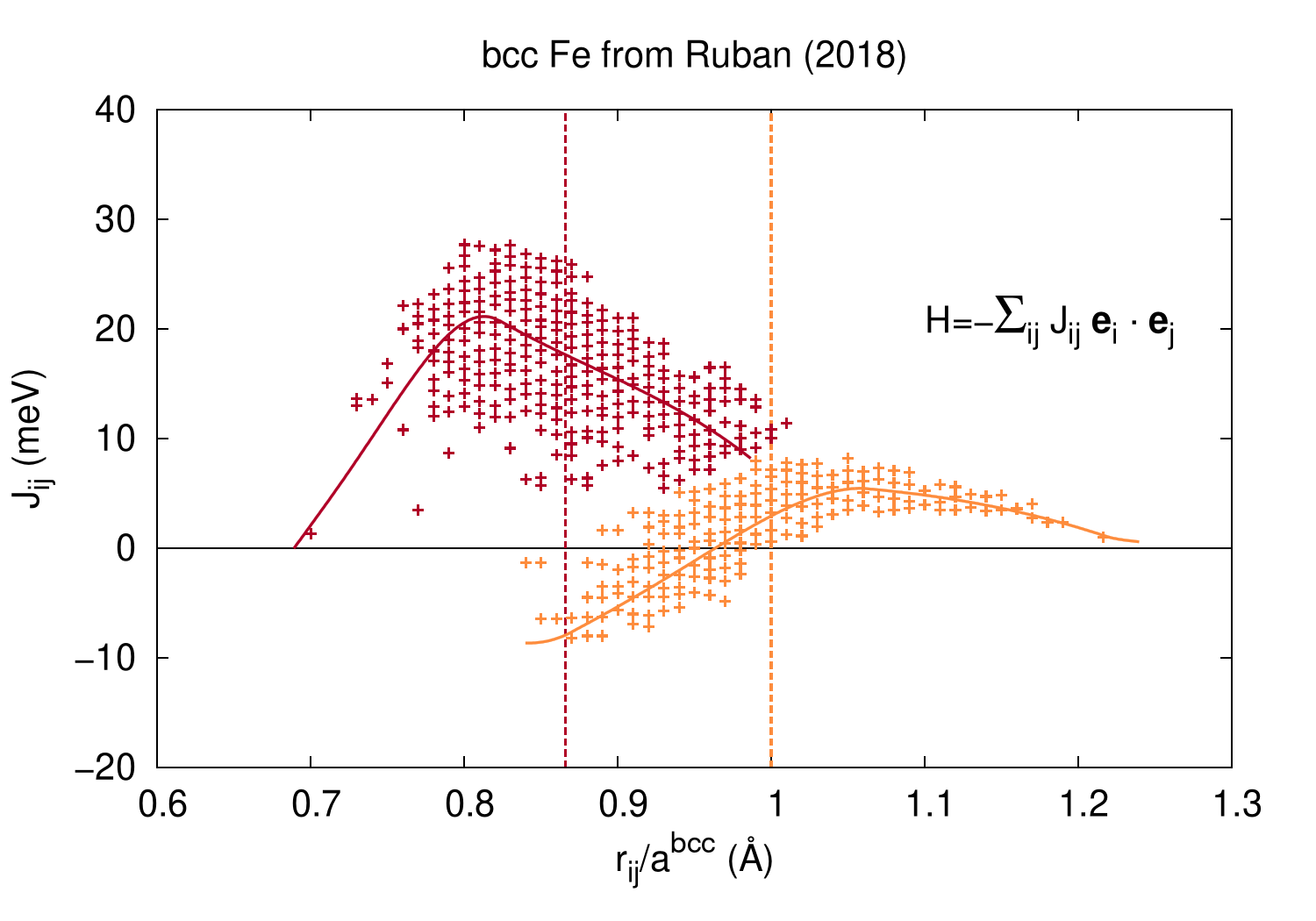}}
\subfigure{\label{fig:bccJijY} \includegraphics[width=0.4\textwidth]{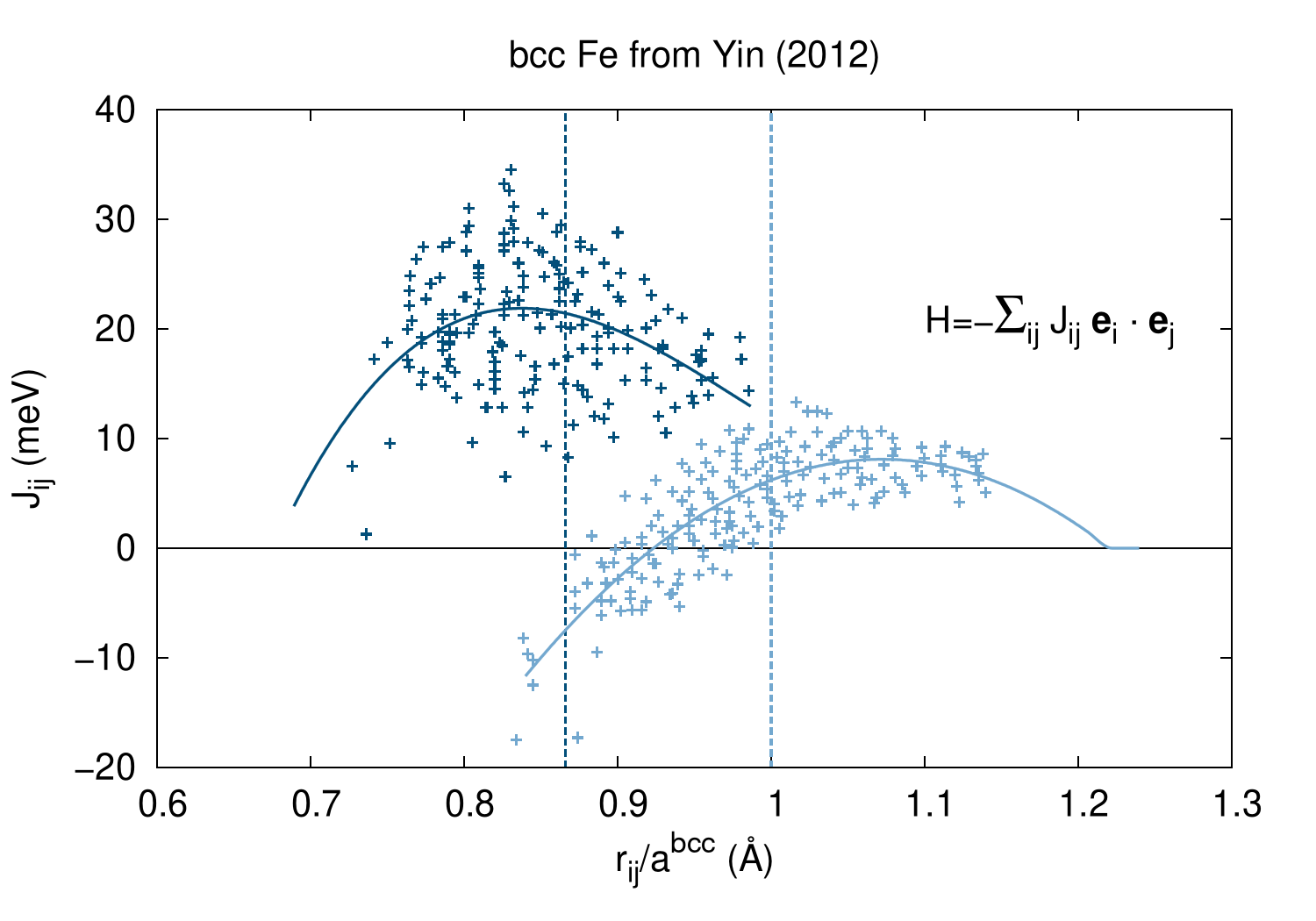}}
\subfigure{\label{fig:fccJij} \includegraphics[width=0.4\textwidth]{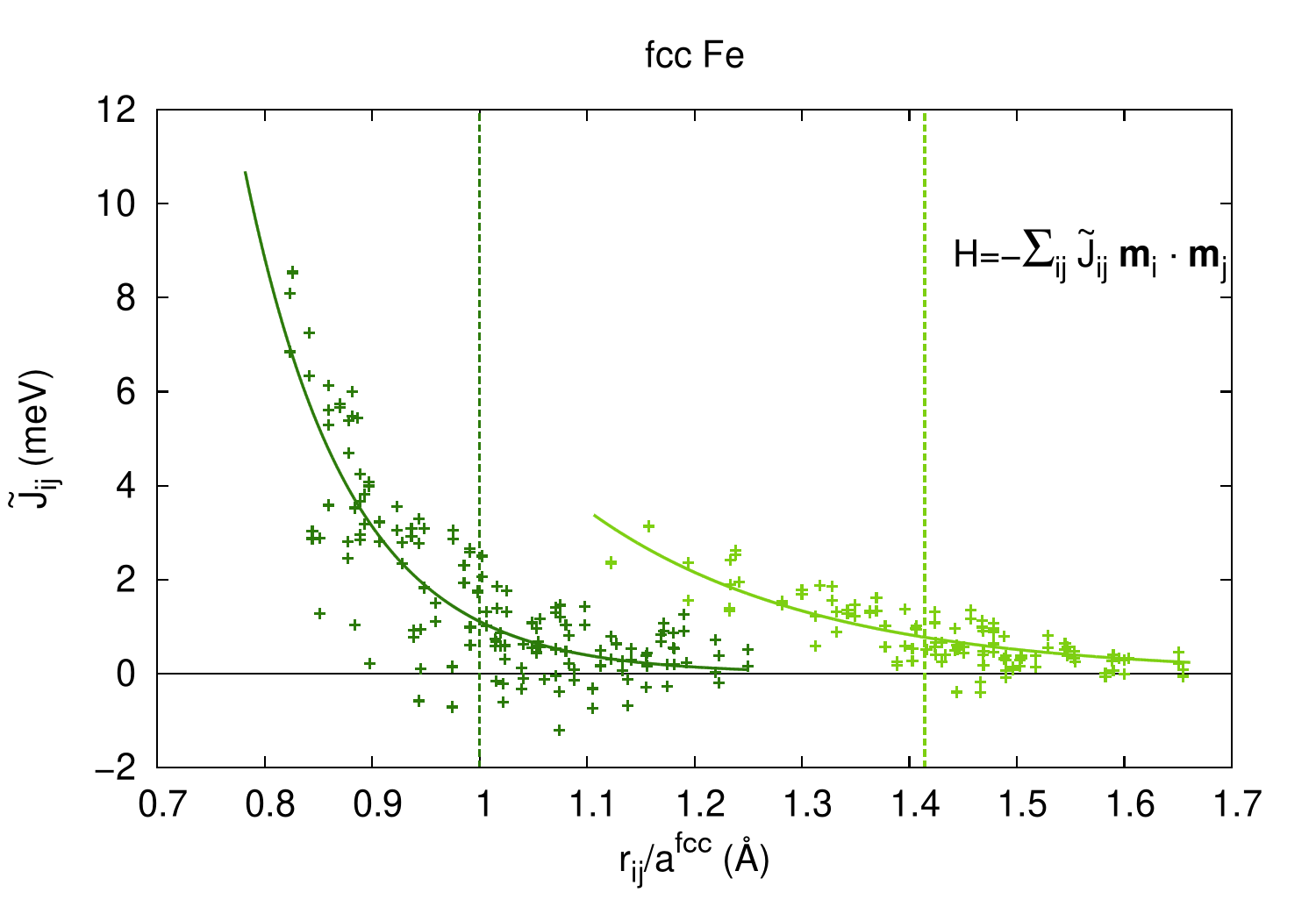}}
\caption{Exchange interactions in in bcc Fe (top and middle) and fcc Fe (bottom) for the first two coordination shells with the parametrizations $J(r_{ij})$ employed in ASD-AIMD simulations. The data points for bcc Fe in the top figure have been extracted from Ref. \cite{RubanJijVib}, in the middle figure from Ref. \cite{YinJijVib}. The vertical dashed lines indicate the ideal neighbors distances for the first two shells. Notice the two different Hamiltonians employed for bcc and fcc Fe. The average size of magnetic moments in fcc Fe is approximately $1.97 \mu_B$. \label{fig:Jij}}
\end{figure}

The pair exchange interactions of bcc Fe on a vibrating lattice have been calculated in Ref.\cite{YinJijVib} and Ref. \cite{RubanJijVib}.
For this system, we consider exchange interactions up to fifth coordination shell, of which for the first two shells we parametrize with polynomial the exchange interactions from Ref. \cite{RubanJijVib} and from Ref. \cite{YinJijVib} as a function of pair distance as shown in Fig. \ref{fig:bccJijR} and Fig. \ref{fig:bccJijY}, respectively. The interactions for the last three shells are considered constant, with values of 0.3, 0.03 and -1.36 meV for third, fourth and fifth coordination shell, respectively.
The exchange interactions for bcc Fe are defined in this work according to the Heisenberg Hamiltonian in the form:
\begin{equation}\label{eq:HeisenbergH}
    H=-\sum_{ij} J_{ij} \textbf{e}_i \cdot \textbf{e}_j,
\end{equation}
where $J_{ij}$ is the exchange interaction between moment $i$ and $j$, and $\textbf{e}_i$ is the unit vector in the direction of moment $i$.
The Curie temperatures derived with the sets of parametrized exchange interactions Ref. \cite{RubanJijVib} and \cite{YinJijVib} are 880 and 1150 K,  respectively, to compare with the experimental value of 1043 K. \cite{TcFe}.
The Curie temperature was calculated with Monte Carlo simulations as a function of temperature using the exchange interactions from the parametrization at ideal lattice pair distances, employing the calculated equilibrium lattice parameter at 1000 K: for the parametrization of the results from Ref. \cite{RubanJijVib}, we employ values of 18.33 and 2.06 meV for the first and second shells, respectively; for the parametrization of the results from Ref. \cite{YinJijVib}, the values are 21.79 and 5.16 meV.

Since we are not aware of any similar investigation of the distance-dependent $J_{ij}$ in fcc Fe, we carried out a DLM-AIMD simulation of this system at temperature $T\approx 1200$ K and volume $V=11.84$  \angstrom$^3$/atom, and we selected two snapshots.
For these snapshots, we drew two random noncollinear configurations of the magnetic moments, and we calculated the exchange interactions with the total energy differences method described in Ref. \cite{Jij_Alling} for several pairs of moments in the first and in the second coordination shells.
The DFT calculations used for the calculation of the exchange interactions were carried out with a $3\times 3\times 3$ k-points mesh  and a cutoff energy of 500 eV.
The resulting exchange interactions  are shown in Fig. \ref{fig:fccJij}.
For fcc Fe, we employ a different form of the Heisenberg Hamiltonian:
\begin{equation}\label{eq:HeisenbergHred}
    H=-\sum_{ij} \Tilde{J}_{ij} \textbf{m}_i \cdot \textbf{m}_j,
\end{equation}
where the full magnetic moments enter explicitly.
For fcc Fe, we employ this definition of the Heisenberg Hamiltonian because the exchange interactions $\Tilde{J}_{ij}=J_{ij}/|\textbf{m}_i||\textbf{m}_j|$,  now independent of the size of the moments, display a smaller scatter of values than with the definition of Heisenberg Hamiltonian of Eq. \ref{eq:HeisenbergH}.
With this Hamiltonian, the scatter would probably be smaller also for bcc Fe, but we do not have access to the sizes of the moments for the calculations in Ref. \cite{YinJijVib} and \cite{RubanJijVib}, therefore we keep the definition of Eq. \ref{eq:HeisenbergH}.
To make easier the comparison of the exchange interactions of fcc and bcc Fe, the size of the magnetic moments in the fcc structure is approximately $1.97 \mu_B$ so that, considering ideal lattice positions, the exchange interactions for first nearest neighbors according to Eq. \ref{eq:HeisenbergHred} are $\Tilde{J}_{ij}=1.1$ meV, the corresponding $J_{ij}$ according to Eq. \ref{eq:HeisenbergH} is of 4.3 meV.
As a comparison, in bcc Fe, the exchange interactions between first nearest neighbors are 17.7 and 21.4 meV for the parametrizations of the exchange interactions from Ref. \cite{RubanJijVib} and Ref. \cite{YinJijVib}, respectively.
For fcc Fe, we consider only the first two coordination shells.

In the present ASD-AIMD simulations, we do not take into account the spread of values but only an average distance-dependence of the exchange interactions, shown in Fig. \ref{fig:Jij}(a-c) by the solid lines passing through the scattered values.
It is clear that these parametrized exchange interactions are not fully representative of the pair interactions in the real systems, but the $J_{ij}$s and the Hamiltonians are employed to govern the spin dynamics, whereas energies are from DFT and therefore include the real interaction between moments beyond the simple distance-dependence.
Any inaccuracies therefore enters only the dynamics of sampling the phase space, not the energetics itself.

However, as will be apparent in Sec. \ref{sec:Results}, the parametrization of exchange interactions plays a central role in the relative stability of the structural phases because different parametrizations sample partly different phase spaces, with possible repercussions on the final results of the present work.
The spin dynamics of a single moment is governed by the effective local field, which is given by the sum over all neighbors of the magnetic moments weighted by their pair exchange interactions, $\textbf{h}_i=\sum_j J_{ij} \textbf{e}_j$.
At the temperatures considered in the present work, around or above the Curie temperature, this local field is small and changes quickly in direction, suggesting that effects of the spread of $J_{ij}$s around our parametrized $J_{ij}(r_{ij})$ values, average out. 
The spin dynamics in the PM state is therefore less sensitive to the details of the instantaneous pair exchange interactions and more to their distance-dependent average, as compared to low temperatures not considered in this work where the background is mainly FM and the local configurations are going to be more important.
From these considerations, we deem reasonable considering only the average distance dependence of the pair exchange interactions.

\section{Thermodynamic path}\label{sec:TIpath}

In the calculation of free energy differences with TI methods, one needs to first define the integration path to employ.
The Gibbs free energy difference between the fcc and bcc structures $\Delta G^{\textrm{fcc-bcc}}(T,P=0)$ can be expressed as:
\begin{equation}
    \begin{split}
        \Delta G^{\textrm{fcc-bcc}}(T_1,P=0)=&\Delta F^{\textrm{fcc}-\textrm{bcc}}(T_1)+\\
        &\Delta F^{\textrm{fcc}}(V^{\textrm{fcc}}_2 \rightarrow V^{\textrm{fcc}}_1,T_1)-\\
        &\Delta F^{\textrm{bcc}}(V^{\textrm{bcc}}_2 \rightarrow V^{\textrm{bcc}}_1,T_1),
    \end{split}
\end{equation}
where the first term, $\Delta F^{\textrm{fcc}-\textrm{bcc}}(T_1)$, is the free energy difference between the fcc and bcc structures at temperature $T_1$ and fixed volumes $V^{\textrm{fcc/bcc}}_2$, and the last two terms in this equation are the free energy differences at constant temperature between two different volumes, $V^{\textrm{fcc/bcc}}_1$ and $V^{\textrm{fcc/bcc}}_2$, which are the equilibrium volumes for each structure at temperatures $T_1$ and $T_2$, respectively.
These latter terms are needed because the calculation of the former term, as will be apparent soon, is carried out at constant volume for different temperatures, therefore needing to re-introduce the equilibrium volume in the calculation in order to account for thermal expansion.
More about the calculation of this term can be found in Sec. \ref{sec:VcorrF}.

The first term, $\Delta F^{\textrm{fcc}-\textrm{bcc}}(T_1)$, can be calculated with TTI \cite{CeriottiTI}, which can be shown to take the following form for free energy differences (see Appendix \ref{app:TTI}):
\begin{equation}\label{eq:TTI}
\begin{split}
    &\Delta F^{\textrm{fcc-bcc}}(T_1)= \Delta F^{\textrm{fcc-bcc}}(T_{\textrm{ref}})\frac{T_1}{T_{\textrm{ref}}}-\\[5pt]
    &T_1\int_{T_{\textrm{ref}}}^{T_1} \frac{\langle E^{\textrm{fcc}} \rangle_{T,V^{\textrm{fcc}}_2} - \langle E^{\textrm{bcc}} \rangle_{T,V^{\textrm{bcc}}_2}}{T^2} \, dT.
\end{split}
\end{equation}
In this equation, the energies $\langle E^{\textrm{s}} \rangle_{T,V^{\textrm{fcc/bcc}}_2}$ at constant volumes $V^{\textrm{fcc/bcc}}_2$ are readily available from the ASD-AIMD simulations, whereas the first term $\Delta F^{\textrm{fcc-bcc}}(T_{\textrm{ref}})$ is a reference free energy difference between the two structures at temperature $T_{\textrm{ref}}$. 
The constant volume employed in the ASD-AIMD simulations from which the energies of Eq. \ref{eq:TTI} was chosen as the equilibrium volume at $T=1000$ K to avoid problems related to diffusion (see Sec. \ref{sec:VcorrF} for a discussion of this problem).

The reference free energy difference is calculated with SSTI (see Sec. \ref{sec:SSTI}). 
The simulations needed to calculate this quantity are carried out in the DLM state, to avoid the need of calculating exchange interactions for the intermediate structures. 
Although the DLM method models an ideal PM state, it is assumed here that it is a good representation of the real magnetic state at very high temperature, near the melting point of Fe.
In this way, the SSTI can be performed at $T_{\textrm{ref}}=1800$ K in the DLM state, and the resulting free energy difference is used directly in \eqnref{eq:TTI}.
The validity of this assumption will be discussed in \secref{sec:Results}.

To summarize, the thermodynamic path employed starts with the calculation of the reference free energy difference with SSTI at $T_{\textrm{ref}}=1800$ K in the DLM state with the 1000 K equilibrium volume, followed by integration over temperature at constant volume in the finite-temperature magnetic state with ASD-AIMD, and finally adding the contribution from thermal expansion/contraction also obtained with ASD-AIMD.

\section{Determination of lattice parameter and free energy contribution from thermal expansion}\label{sec:VcorrF}

The equilibrium volume as a function of temperature for each structure and exchange interaction parametrization is determined by running ASD-AIMD simulations at different volumes for a few selected temperatures, and then interpolating the pressure as a function of volume at constant temperature with a second order polynomial.
The equilibrium volume is the volume at zero pressure.

For bcc Fe at 1800 K around the equilibrium volume, we observed some concerted diffusion events as previously reported for bcc Fe at high pressures and temperatures \cite{DiffusionFeHTHP}, as well as bcc Ti \cite{DiffusionbccTi} and other bcc metals \cite{DiffusionbccMetals} at ambient pressures and temperatures close to melting.
It is not clear if these events are a real effect or just an artifact originated from the low accuracy parameters employed, since the pressure is strongly underestimated as a result of the parameters employed in these simulations; additional investigations are needed to address the origin of this effect.
For what concerns the present investigation, avoiding diffusion is desirable since it makes more difficult to converge quantities, requiring much longer simulation times: we therefore perform all the simulations at the equilibrium volume at $T_0=1000$ K to hinder diffusion, and the TI is carried out on results from these simulations.
Diffusion could be important to stabilize a certain crystal structure at high temperature \cite{SSTIKlarbring,DiffusionFeHTHP,DiffusionbccTi}, but from the low frequency of these events we deem not fundamental the inclusion of this effect in the investigation of phase stability of Fe.

\begin{figure}%
\centering
 \includegraphics[width=0.45\textwidth]{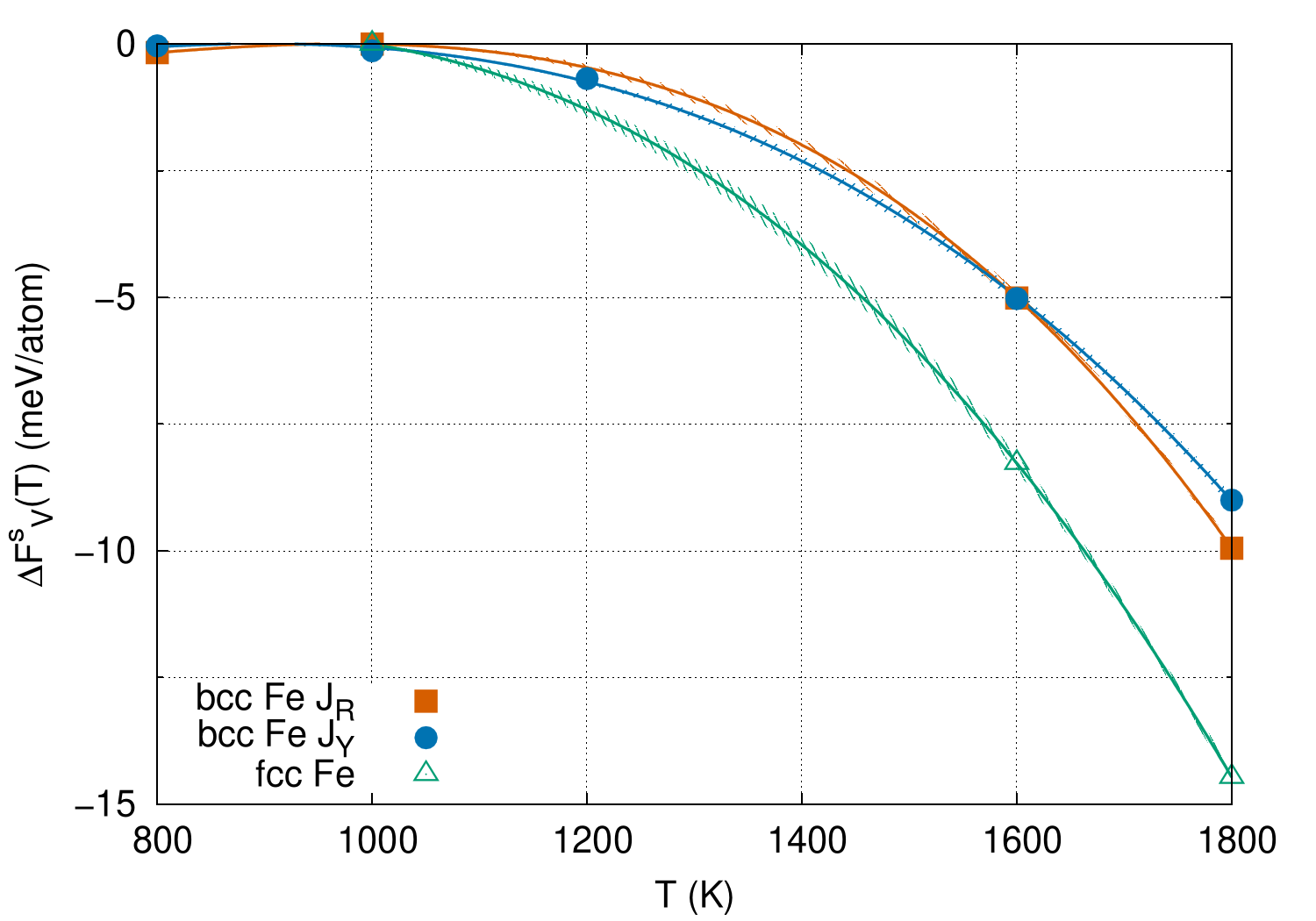}
\caption{Contribution to the free energy due to thermal expansion as a function of temperature for bcc Fe with exchange interactions parametrized after Ref. \cite{RubanJijVib} ($\textrm{J}_\textrm{R}$, red squares and curve), bcc Fe with exchange interactions parametrized after Ref. \cite{YinJijVib} ($\textrm{J}_\textrm{Y}$, blue circles and curve), and for fcc Fe (green triangle and curve). The curves are a cubic polynomial interpolation to the data points, calculated with Eq. \ref{eq:VcorrF}. \label{fig:dFVcorr}}
\end{figure}

Since the simulations are carried out at all temperatures fixing the volume to the 1000 K equilibrium value, in order to obtain the Gibbs free energy at $p=0$ GPa, one needs to calculate the Helmholtz free energy contribution due to thermal expansion for each structure $\Delta F^{\textrm{s}}(V_0 \rightarrow V_1,T_1)$ from volume $V_0=V_{\textrm{eq}}(T_0)$, with $T_0=1000$ K, to volume $V_1=V_{\textrm{eq}}(T_1)$.
We calculate this contribution for each structure and parametrization of the exchange interactions at a few different temperatures $T_1$ by integrating the pressure at temperature $T_1$:
\begin{equation}\label{eq:VcorrF}
    \Delta F(V_0 \rightarrow V_1,T_1)=-\int_{V_0}^{V_1} P(V,T_1)dV,
\end{equation}
where $P(V,T_1)$ is the quadratic interpolation of the pressure previously mentioned in this section.
The free energy contribution for each structure and parametrization is shown in Fig. \ref{fig:dFVcorr}, where the lines correspond to a fitting with a third-order polynomial.

\section{Stress-strain thermodynamic integration (SSTI)}
\label{sec:SSTI}

% State main equations and show that it works at 0 K even if the volumes are not the 0-pressure volumes (\figref{fig:SSTI0K}).

The SSTI method consists in integrating the thermodynamic stresses in the cell along a deformation path between two structures \cite{Wallace,SSTI_example_III,SSTIKlarbring}.
The deformation path is described with a matrix made of the lattice vectors of the intermediate structures, similarly to the matrices defined in \eqnref{eq:bccfccmatrices}, which depends on a parameter $\lambda$ that brings continuously from the initial to the final structure.
In this case, we define the deformation path as a linear interpolation between the bcc and fcc matrices, namely:
\begin{equation}\label{eq:hlambdamatrix}
    \textbf{h}(\lambda)= \textbf{h}^{\textrm{bcc}}+\lambda ( \textbf{h}^{\textrm{fcc}}-\textbf{h}^{\textrm{bcc}}). 
\end{equation}
For the bcc to fcc transformation, this is similar to the well known Bain path, which consists simply in changing the $c/a$ ration in the bcc cell.
With the definition of deformation path of Eq. \ref{eq:hlambdamatrix}, the free energy difference between the two structures can be shown to be \cite{SSTIKlarbring,SSTI_example_III}:
\begin{equation}\label{eq:SSTIequation}
    \Delta F^{\textrm{fcc-bcc}}=\int_0^1 V(\lambda) \pmb{\sigma}(\lambda)\textbf{h}^{-T}(\lambda):(\textbf{h}^{\textrm{fcc}}-\textbf{h}^{\textrm{bcc}}) \textrm{d}\lambda, 
\end{equation}
where $V(\lambda)$ is the volume of the configuration with parameter $\lambda$, $\pmb{\sigma}(\lambda)$ the Cauchy stress tensor for this configuration, $^{-T}$ indicates the inverse and transposed matrix, and finally $:$ indicates a contraction of the two matrices over both indices, $i.e.$, $\textbf{A}:\textbf{B}=\sum_{i,j}A_{ij}B_{ij}$.
The calculation of the free energy difference with this method requires to collect the stress tensor for the initial structure, the final structure, and a certain number of intermediate structures, by carrying out dynamic simulations.
The SSTI method has been previously used in simplified forms \cite{SSTI_example_I,SSTI_example_II} and with the full equation \cite{SSTI_example_III,SSTIKlarbring}, but up to our knowledge it has never been employed accounting explicitly for the magnetic degrees of freedom.

\begin{figure}%
\centering
\subfigure{\label{fig:StressSSTI0K} \includegraphics[width=0.45\textwidth]{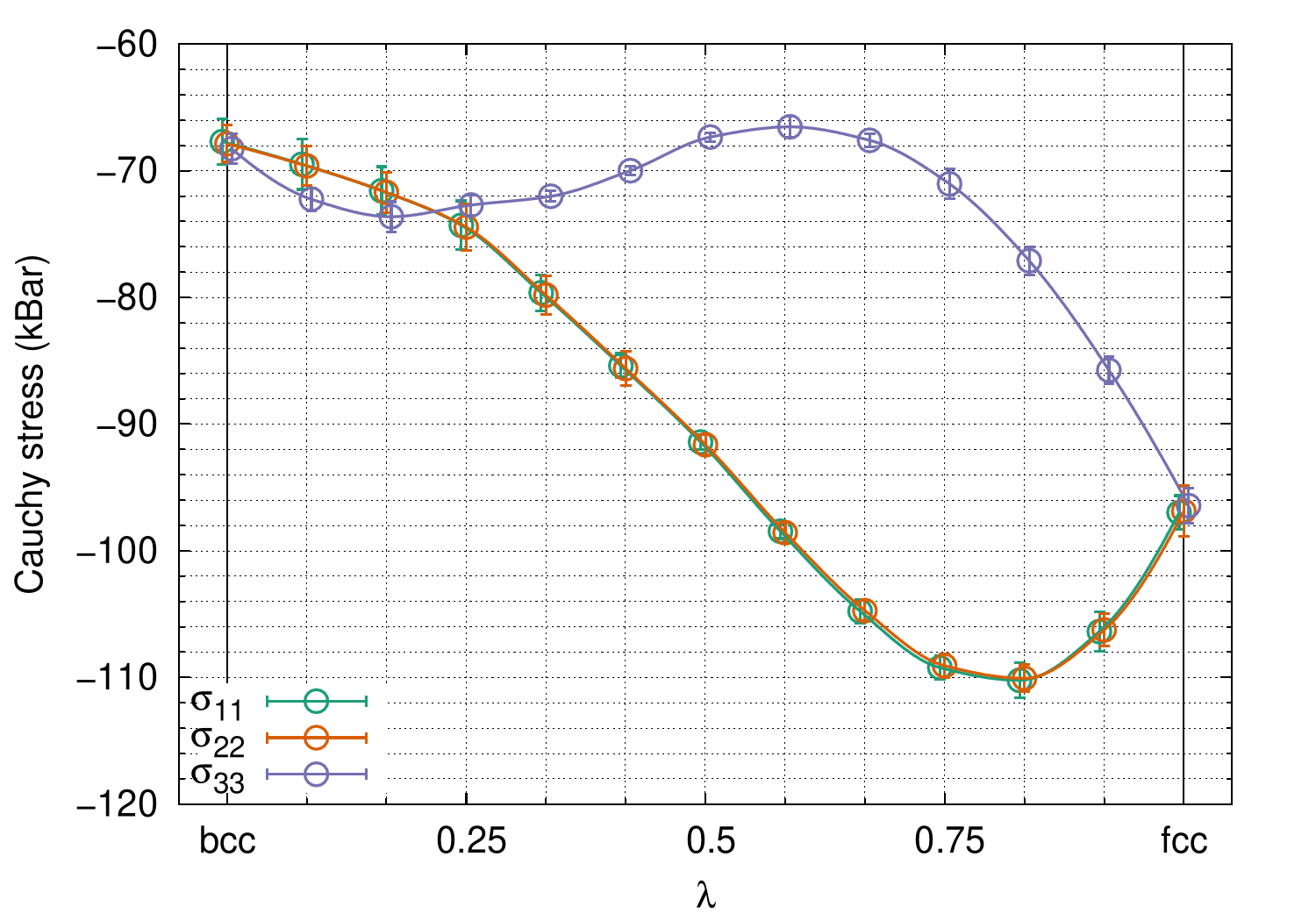}}
\subfigure{\label{fig:DeltaFSSTI0K} \includegraphics[width=0.45\textwidth]{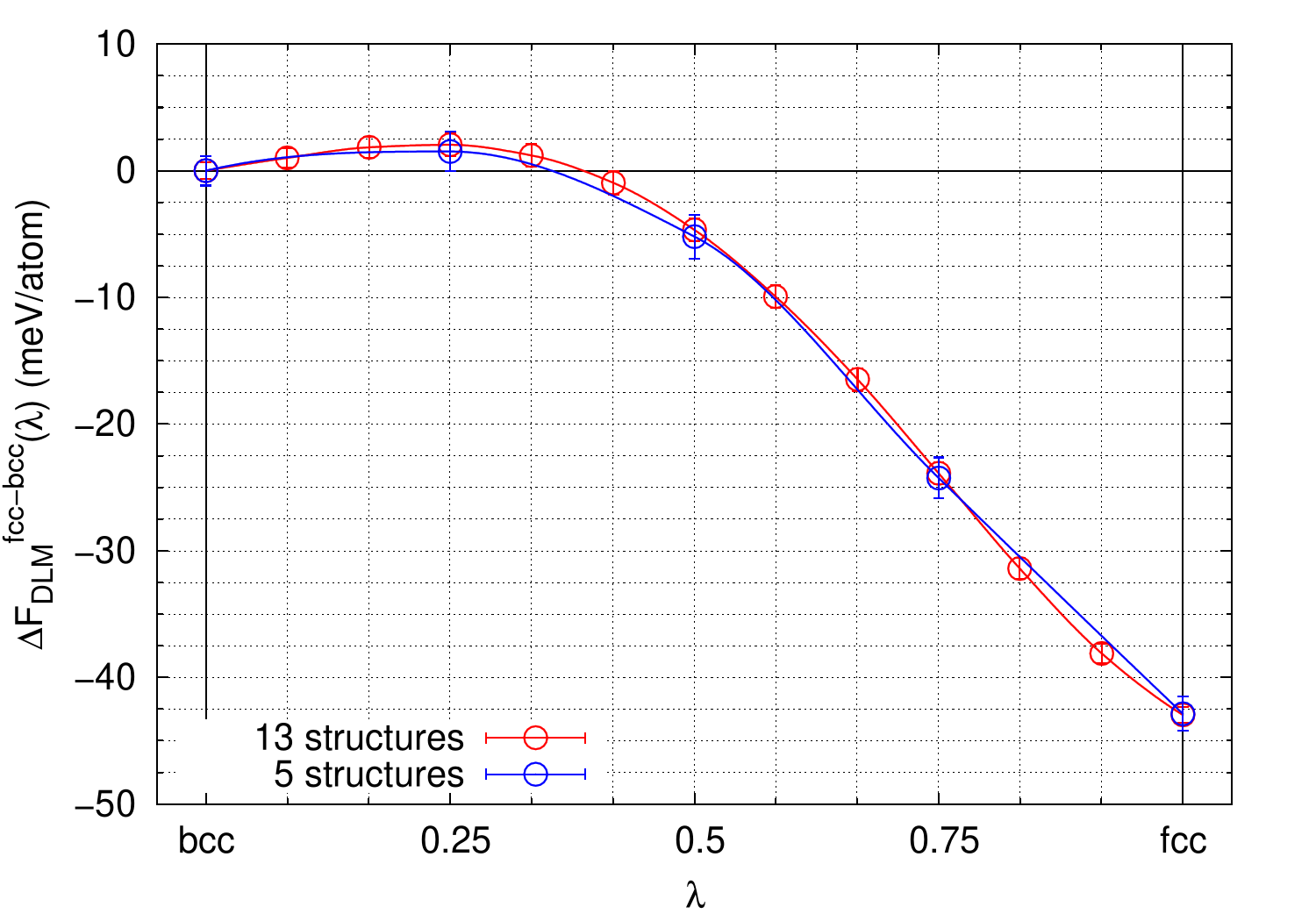}}
\caption{Isotropic components of the stress tensor $\sigma_{11}$, $\sigma_{22}$ and $\sigma_{33}$ (top) and energy difference with respect to bcc Fe calculated with SSTI (bottom) along the deformation path at 0 K in the DLM state. Isotropic components are the only components contributing to the energy difference. The final energy difference ($-43.0 \pm 0.3$ meV/atom) coincides, within statistical error, with the direct energy difference between fcc and bcc ($-42.9 \pm 0.4$ meV/atom).  \label{fig:SSTI0K}}
\end{figure}

Since the DLM state is employed, we check the validity of the method by carrying out a SSTI calculation  at $T=0$ K in the DLM state of the free energy difference, which is just the energy difference.
To describe the DLM state in this test, six magnetic configurations, generated with Monte Carlo simulations such that the spin-spin correlation functions are as close as possible to zero, are employed for all structures.
We use eleven intermediate structures between bcc and fcc and interpolate the integrand of \eqnref{eq:SSTIequation} with piecewise cubic Hermite polynomials.
The isotropic components of the stress tensor along the deformation path are shown in \figref{fig:StressSSTI0K}, and the resulting energy difference along the path in \figref{fig:DeltaFSSTI0K}.
Only the isotropic components of the stress tensor are presented because it can be shown that these are the only components contributing to the integral in this case.
The energy difference between fcc and bcc Fe obtained with the SSTI method ($-43.0 \pm 0.3$ meV/atom) and from direct difference ($-42.9 \pm 0.4$ meV/atom) are virtually identical.
Since at finite temperatures we cannot run DLM-AIMD simulations for 11 intermediate structures for computational reasons, we check how the method works employing only three intermediate structures equally spaced from each other (blue circles and lines in \figref{fig:DeltaFSSTI0K}), and we observe a negligible difference in the final result ($\approx 0.1$ meV/atom).
This test suggests that the method is reliable in the magnetically disordered state.

%\section{Results}
%\label{sec:ResultsSSTI+mTPT}
\section{Results} \label{sec:Results}

\begin{figure}
    \centering
    \includegraphics[width=0.45\textwidth]{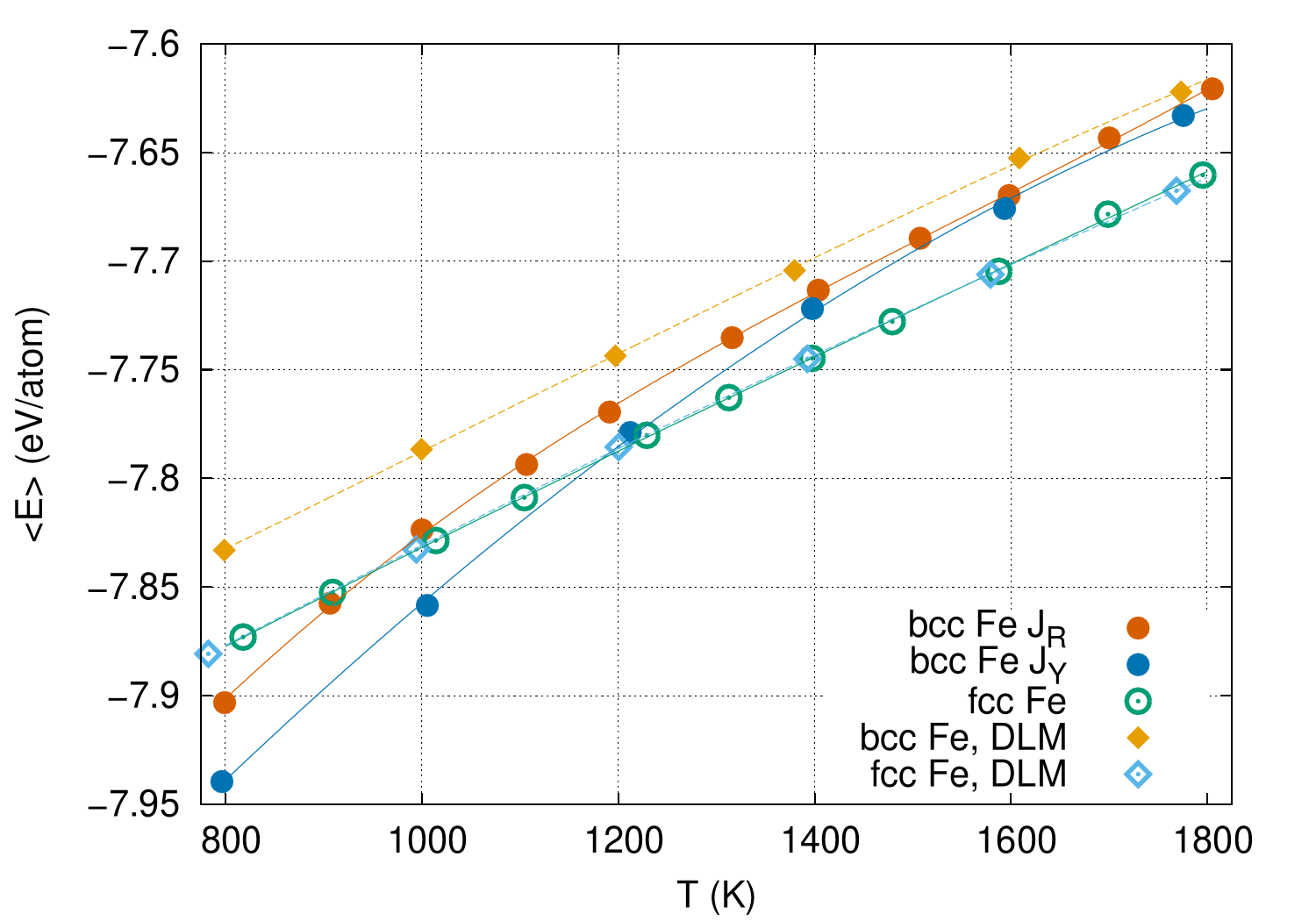}
    \caption{Total energy as a function of temperature of bcc Fe with parametrized exchange interactions from Ref. \cite{RubanJijVib} (red), bcc Fe with parametrized exchange interactions from Ref. \cite{YinJijVib} (blue) and fcc Fe (green) from ASD-AIMD (solid lines and circles), and bcc Fe (yellow) and fcc Fe (light blue) from DLM-AIMD (dashed lines and empty diamonds) simulations. Error bars are smaller than the symbols.}
    \label{fig:EvsTSDMDDLMMD}
\end{figure}

The total energies as a function of temperature obtained from ASD-AIMD and DLM-AIMD are shown in \figref{fig:EvsTSDMDDLMMD}.
First of all, we notice that the total energy of fcc Fe from ASD-AIMD and DLM-AIMD are practically indistinguishable along the whole temperature range, demonstrating that fcc Fe is well in the PM state at these temperatures.
For bcc Fe, on the contrary, the total energy is much lower at low temperatures ($\approx 100$ meV/atom lower) when magnetic interactions are considered as compared to the DLM state.
At high temperatures, however, the total energy for both parametrizations of exchange interactions of bcc Fe approach the DLM limit, with the stronger exchange interactions from Ref. \cite{YinJijVib} leading to a slightly lower energy at $T=1800$ K than the DLM energy.

\begin{figure}
    \centering
    \includegraphics[width=0.45\textwidth]{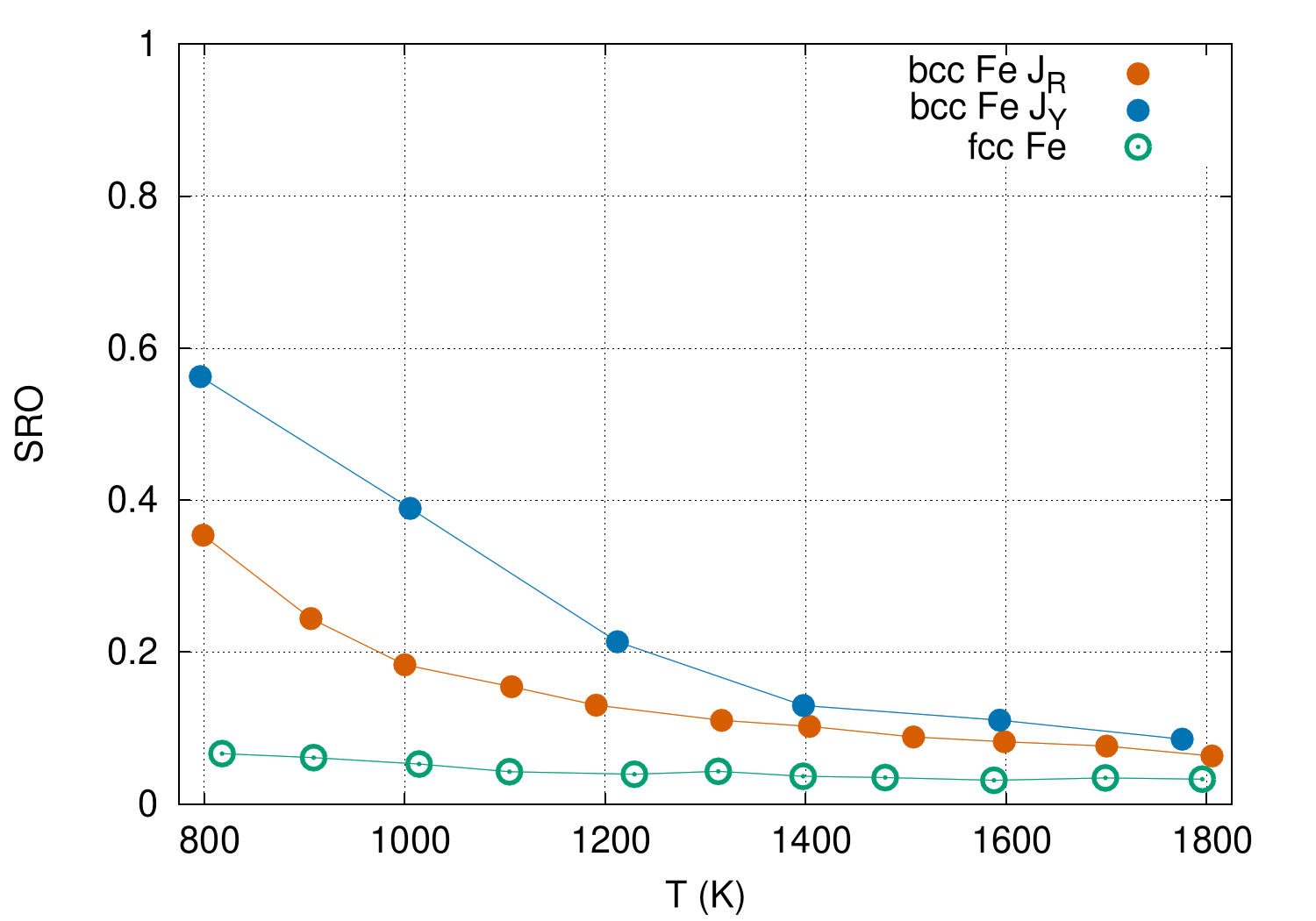}
    \caption{SRO for first nearest neighbors as a function of temperature of bcc Fe with parametrized exchange interactions from Ref. \cite{RubanJijVib} (red), bcc Fe with parametrized exchange interactions from Ref. \cite{YinJijVib} (blue) and fcc Fe (green) from ASD-AIMD. SRO=0 corresponds to DLM state, whereas SRO=1 corresponds to pure FM state. The lines are just a guide to the eye.}
    \label{fig:SROvsT}
\end{figure}

In order to see more in detail how different the finite-temperatures magnetic state and the DLM state are, the magnetic short range order (SRO) parameter for first nearest neighbors from ASD-AIMD simulations is shown for bcc Fe with the two parametrizations and for fcc Fe in \figref{fig:SROvsT}. 
Of course, the DLM state has zero SRO, since it models an ideal PM state.
A certain degree of SRO survives at the highest temperature even for fcc Fe, however this does not seem to particularly affect the energies, as seen in Fig. \ref{fig:EvsTSDMDDLMMD}. 
From inspection of energies and SRO, we can therefore imagine that using the free energy difference between fcc and bcc structures calculated in the DLM state at $T=1800$ K in \eqnref{eq:TTI} is going to introduce a small error; however, the energy difference between the different magnetic states of $\approx 1$ meV/atom will cancel out with the difference in magnetic entropy, resulting in an error below our statistical resolution.

% Show final results, comment how better Js in bcc should improve the results. Plot on comparison between my results, Dudarev's, Fritz's and CALPHAD????
% Predicted transition temperatures for $T_{\textrm{ref}}=$1800 K: $T^{\alpha \xrightarrow{} \gamma} \approx $ 830 K; $T^{\gamma \xrightarrow{} \delta} \approx $ 1450 K. Minimum Gibbs free energy difference: -3 meV/atom at $T\approx$ 1050 K.

%\subsection{Comparison between SSTI+mTPT and TTI free energy differences} \label{sec:ResultsSSTIvsTTI}

\begin{figure}%
\centering
\subfigure{\label{fig:StressSSTIvsT} \includegraphics[width=0.45\textwidth]{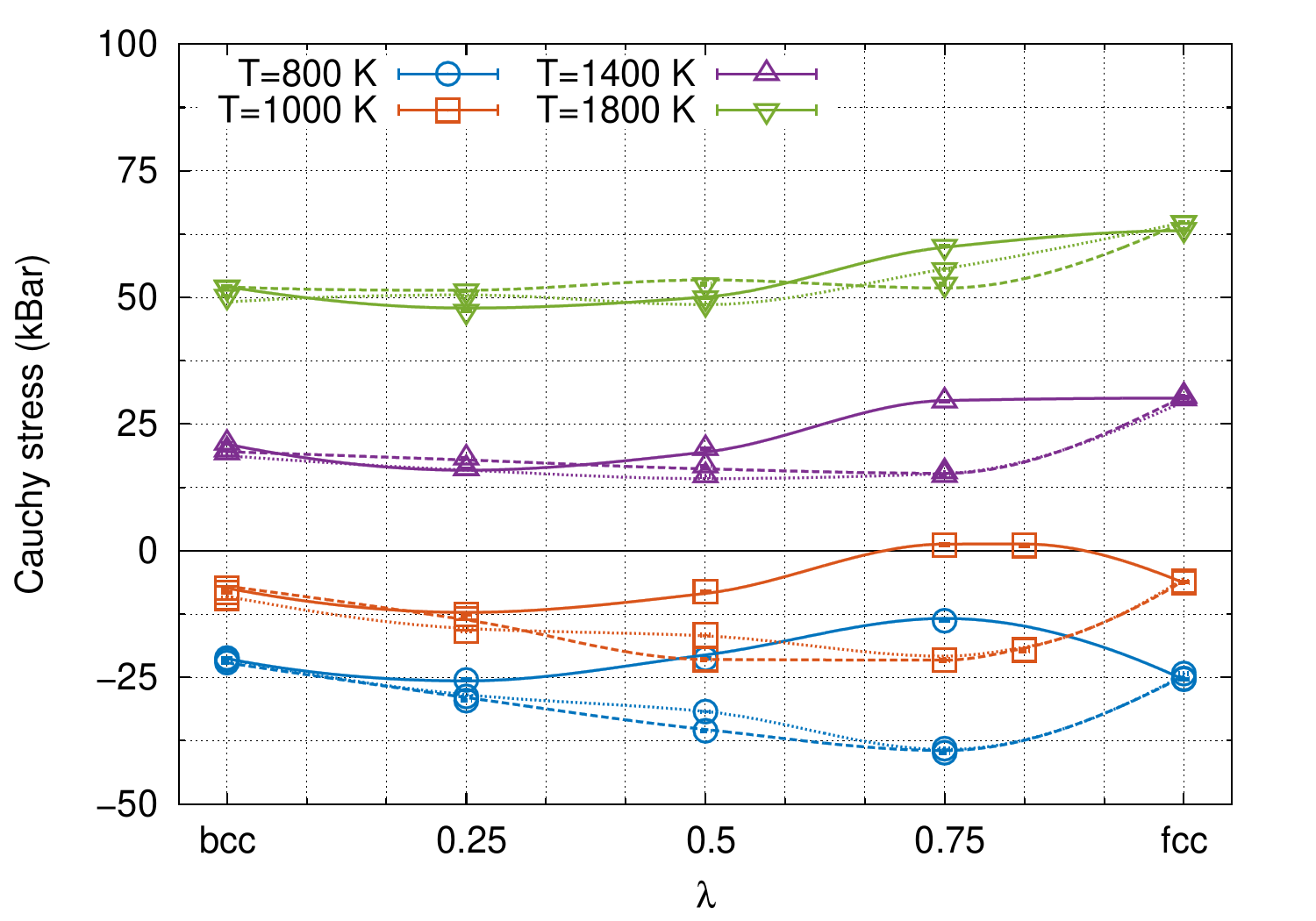}}
\subfigure{\label{fig:DeltaFSSTIvsT} \includegraphics[width=0.45\textwidth]{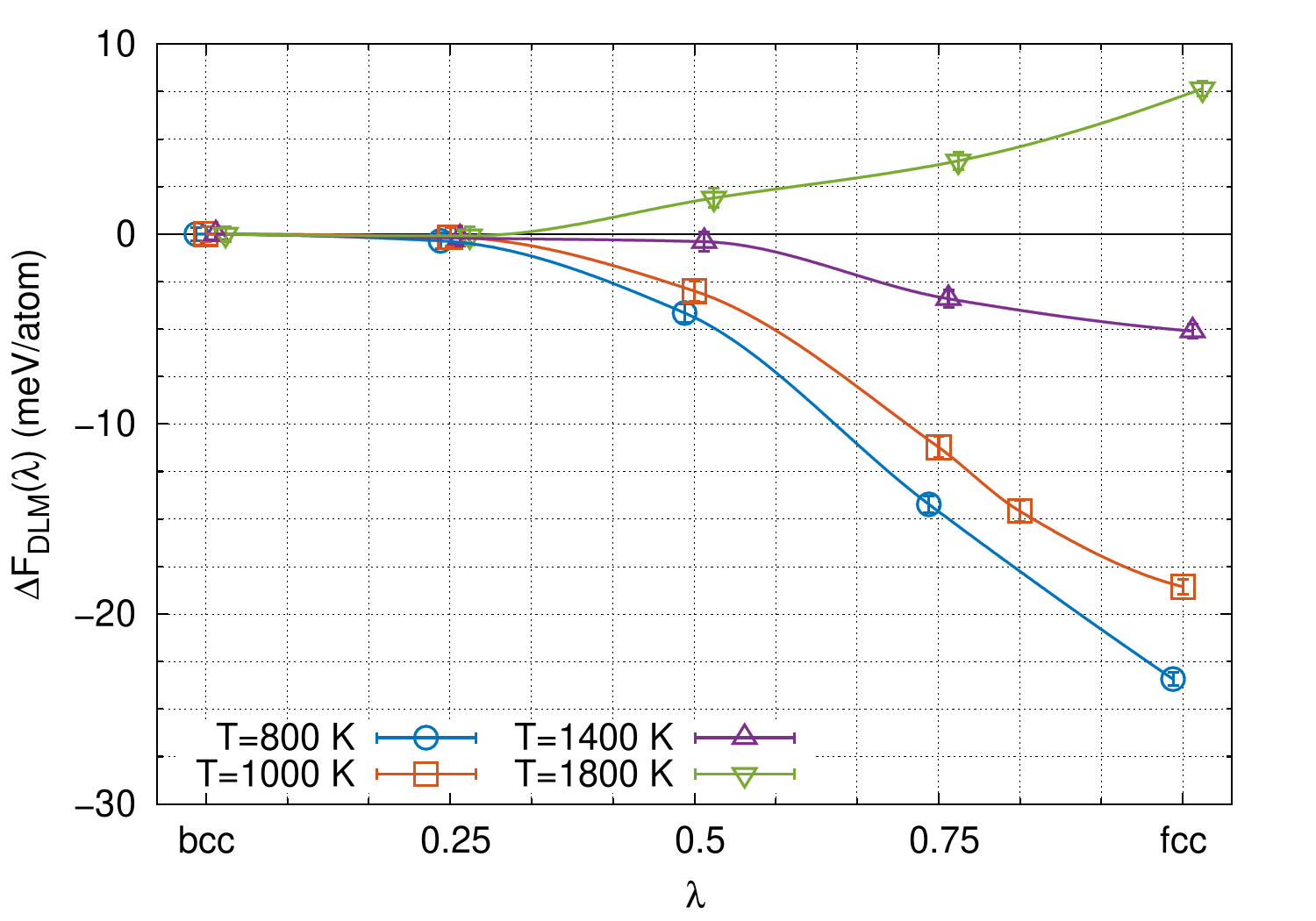}}
\caption{(Top) isotropic components of the stress tensor $\sigma_{11}$ (dotted line), $\sigma_{22}$ (dashed line) and $\sigma_{33}$ (solid line) and (bottom) free energy difference with respect to bcc Fe calculated with SSTI in the DLM state along the deformation path for the different temperatures.
\label{fig:SSTIvsT}}
\end{figure}

Having established that the DLM and the finite-temperature magnetic state are similar enough at high temperatures, we can pass to the results of the SSTI procedure to obtain free energy differences between bcc and fcc Fe in the DLM state.
We have carried out the calculation of $\Delta F^{\textrm{fcc-bcc}}_{\textrm{DLM}}$ at temperatures $T= 800, 1000, 1400$, and 1800 K to see how the stresses along the deformation path are affected by temperature. 
In addition, we obtain from here the temperature dependence of the free energy difference in the DLM state so that we are able to address the effect of magnetic interactions and SRO on the structural transitions.
The diagonal components of the stress tensor and free energy differences from SSTI for the different temperatures at constant volume in the DLM state are shown in \figref{fig:SSTIvsT}.
All the stresses, obviously, increase with temperature; in addition, we observe that the difference between the $\sigma_{33}$ component (the highest in each curve) and the other two components along the deformation path decreases with increasing temperature.
At 1800 K, the stresses become fairly isotropic for every intermediate structure, suggesting that the systems are close to melting.
We did observe, indeed, cases of concerted diffusion in these intermediate structures at the highest temperature, but it is difficult to estimate if the diffusion is due to the low accuracy parameter employed in running the simulations or it is a real physical effect, similarly to what is observed in bcc Fe (\secref{sec:VcorrF}).

The free energy difference from bcc Fe along the path shows that, in the DLM state, the bcc structure and the first intermediate structure at finite temperatures are fairly similar in free energy.
This does not mean that this intermediate structure is as stable as the bcc structure, since here we consider only a fixed volume, which is not the equilibrium volume for any structure in the DLM state. 
Nonetheless, it is interesting that bcc and this intermediate, bct structure are similar, having possible implications for the martensitic phase of steels.

\begin{figure}
    \centering
    \includegraphics[width=0.45\textwidth]{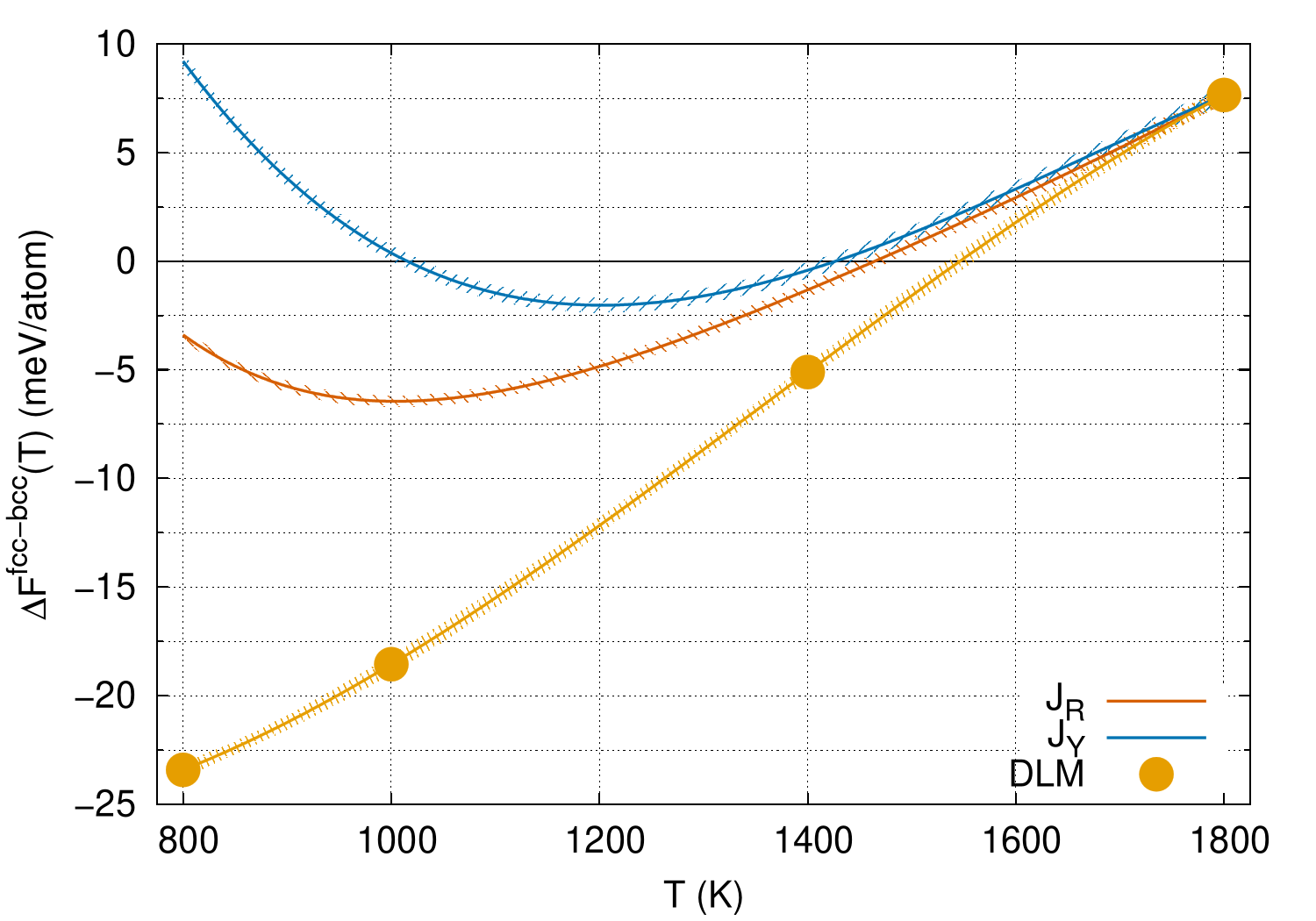}
    \caption{The free energy difference between DLM fcc and DLM bcc Fe at constant volume $\Delta F^{\textrm{fcc-bcc}}(T)$ as a function of temperature (yellow circles and lines) and in the finite-temperature magnetic states with exchange interactions for bcc Fe parametrized after Ref. \cite{RubanJijVib} ($\textrm{J}_{\textrm{R}}$, red line) and after Ref. \cite{YinJijVib} ($\textrm{J}_{\textrm{Y}}$, blue line) at the 1000 K equilibrium volumes.
    }
    \label{fig:dFconstantV}
\end{figure}

The free energy difference in the DLM state obtained with SSTI is shown in \figref{fig:dFconstantV} (yellow circles and lines), where negative values indicate fcc stability and positive values indicate bcc stability.
Clearly, in the DLM state, the fcc structure is more stable than the bcc at low temperatures, and at higher temperatures the larger vibrational entropy related to the openness of the bcc structure comes into play and makes the phase transition possible.
Of course, since the volume is kept constant at all temperatures, we cannot exclude that a contribution from thermal expansion would change the relative stability.
Nonetheless, to elucidate the effect of magnetic order and vibrations on the structural transitions in Fe, we can calculate the free energy difference at constant volume in the finite-temperature magnetic states with TTI, using as reference free energy the value from SSTI at $T=1800$ K (blue and red lines in \figref{fig:dFconstantV}).
As it can be seen, the double transition $\alpha \rightarrow \gamma \rightarrow \delta$ occurs even at constant volume in the parametrization of exchange interactions for bcc Fe from Ref. \cite{YinJijVib}, and it can be expected to occur at $T<800$ K for the parametrization of exchange interactions for bcc Fe from Ref. \cite{RubanJijVib}.
These results show that the FM ordering of bcc Fe makes this phase the thermodynamically stable phase at low temperatures due to lower energy compared to the fcc structure.
The transition then occurs thanks to a combination of increased magnetic energy of the bcc phase and, at the same time, a higher magnetic entropy of the fcc phase.
This latter information can be inferred from the SRO parameter (Fig. \ref{fig:SROvsT}), where it is clear that in the fcc phase magnetic moments are more free to rotate than in the bcc phase.

\begin{figure}
    \centering
    \includegraphics[width=0.5\textwidth]{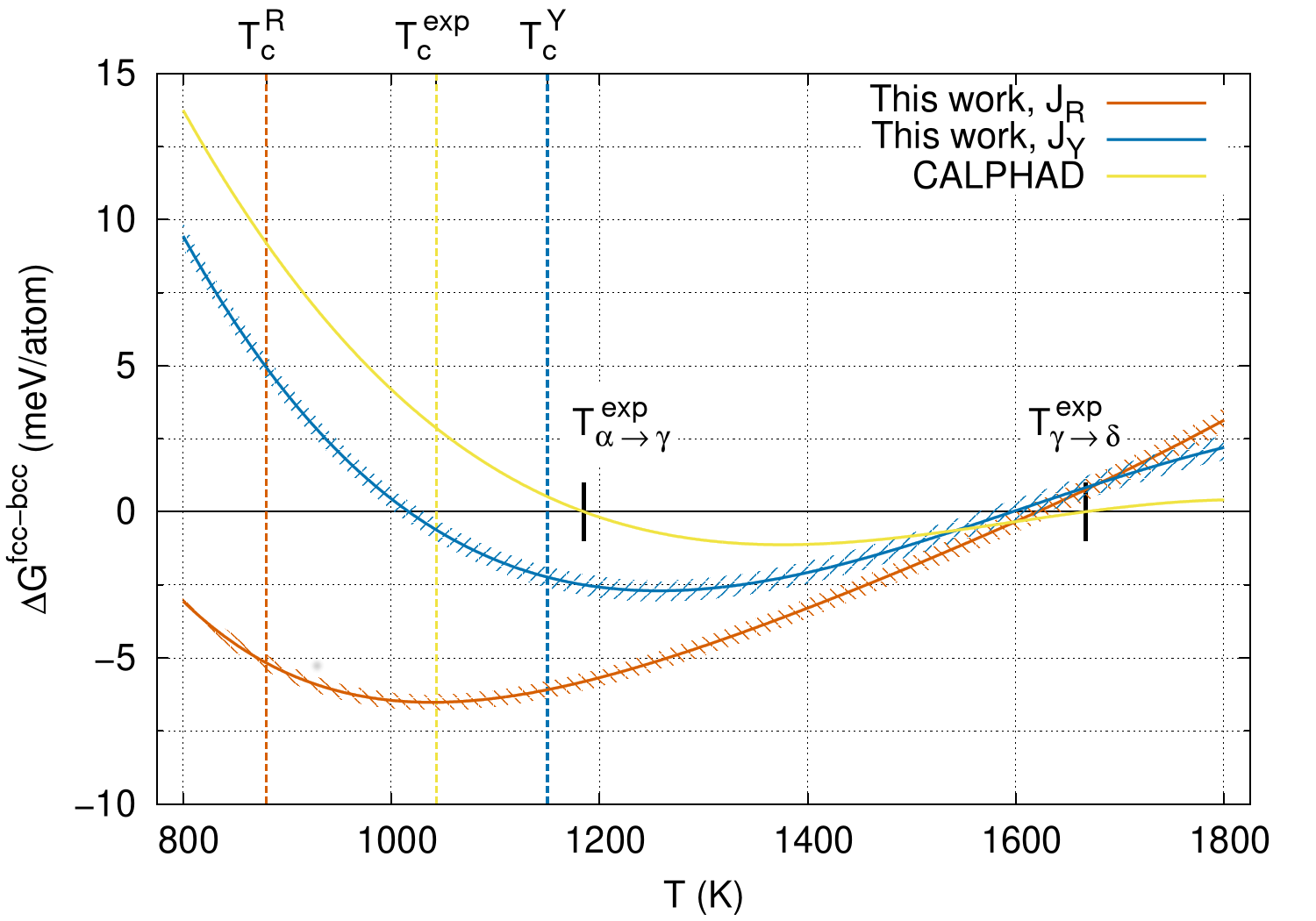}
    \caption{Gibbs free energy difference as a function of temperature from the present work with exchange interactions for bcc Fe parametrized after Ref. \cite{RubanJijVib} ($\textrm{J}_{\textrm{R}}$, red), with exchange interactions for bcc Fe parametrized after Ref. \cite{YinJijVib} ($\textrm{J}_{\textrm{Y}}$, blue), and from CALPHAD (yellow line) \cite{FritzThesis}. The shaded areas indicate error bars corresponding to twice the standard error. The vertical lines indicate the Curie temperature of bcc Fe in the corresponding model (CALPHAD indicates experimental).}
    \label{fig:DeltaGvsLiterature}
\end{figure}

Finally, the full Gibbs free energy difference between fcc and bcc Fe at zero pressure including thermal expansion is shown in \figref{fig:DeltaGvsLiterature} for both parametrizations of exchange interactions of bcc Fe employed, together with results from the CALPHAD method (values digitally extracted from Ref. \cite{FritzThesis}) which can be thought as the experimental counterpart.
First and foremost, these results show that, no matter the parametrization of exchange interactions employed for bcc Fe, the Gibbs free energy difference curves show a minimum below zero, indicating that the double transition occurs in both cases: in the case of the parametrization after Ref. \cite{YinJijVib}, the two transitions occur in the investigated temperature range, in the case of the parametrization after Ref. \cite{RubanJijVib}, the $\alpha \rightarrow \gamma$ transition occurs below $T=800$ K.
The difference between the two curves calculated in this work also reveals the importance of exchange interactions and the Curie temperature of bcc Fe on the lower structural transition: the lower the Curie temperature is (shown for both parametrizations in \figref{fig:DeltaGvsLiterature} as  dashed, vertical lines), the lower the $\alpha \rightarrow \gamma$ transition occurs.
The dependence of the structural transition on Curie temperature is partly an artifact of the present scheme,  where the exchange interactions for bcc Fe are taken from literature, since the exchange interactions and therefore the Curie temperature are strictly connected to the energetics of the system.
Nonetheless, inclusion of thermal expansion does not change the picture compared to Fig. \ref{fig:dFconstantV} for the $\alpha \rightarrow \gamma$ transition, since the volumes employed in SSTI for bcc and fcc Fe are the equilibrium volumes at 1000 K.

The $\gamma \rightarrow \delta$ transition is well reproduced (within 50 K) independently of the exchange interactions employed, stressing how the details of magnetic ordering do not play any particular role at these temperatures and the important physics is related mainly to the vibrational DOF within a magnetically disordered state.
Despite the remarkable agreement between our simulations and CALPHAD results, we do also observe a qualitative difference between the present calculations and experiments: in both calculated curves, the alpha to gamma transition temperature is lower than the Curie temperature of bcc Fe in the corresponding model, in contrast with experiments.

As a final remark, we notice that our calculated Gibbs free energy difference between bcc and fcc Fe calculated using the exchange interactions for bcc Fe from Ref. \cite{YinJijVib} is within 5 meV/atom from the CALPHAD estimate.
This small difference is enough to induce an error of $\approx 150 K$ in the predicted structural transition at low temperature, demonstrating the unforgiving nature of Fe as a benchmark system for modelling of phase stability.

\section{Conclusions} \label{sec:Conclusions}

In this work, the phase stability in Fe at zero pressure was investigated from first principles with thermodynamic integration methods based on ASD-AIMD simulations.
The Gibbs free energy difference between fcc and bcc Fe as a function of temperatures is calculated starting from a representation of an ideal paramagnetic state with the DLM approach, which is then reconnected to finite-temperature magnetism. 
A thermodynamic integration over stress-strain variables is carried out at $T=1800$ K to calculate the free energy difference in the DLM state with 
DLM-AIMD simulations, and it is then used as a reference free energy difference in thermodynamic integration over temperature based on
coupled ASD-AIMD simulations, which enable to simulate vibrational and magnetic degrees of freedom consistently.
To run these simulations, we calculated the exchange interactions as a function of pair distance in fcc Fe in presence of vibrations, and we employed two different parametrization of the exchange interactions for bcc Fe from the literature.

The $\alpha \rightarrow \gamma \rightarrow \delta$ transitions in Fe at ambient pressure are correctly predicted by the present method for both parametrization of the bcc exchange interactions, and for the $\gamma \rightarrow \delta$ transition the predicted transition temperature is within $\approx 50$ K from experimental values.
The present work devises a thermodynamic path that enables the calculation of Gibbs free energy differences in magnetic materials from first principles approaching the long-sought for accuracy in the order of 1 meV/atom.

\hfill \break

\begin{acknowledgments}
\noindent
The computations were enabled by resources provided by the Swedish National Infrastructure for Computing (SNIC) located at National Super Computer Centre (NSC) in Link\"oping, partially funded by the Swedish Research Council through Grant Agreement No. 2018-05973. 
B.A. acknowledges financial support from the Swedish Research Council (VR) through Grant No. 2019-05403, from the Swedish Government Strategic Research Area in Materials Science on Functional Materials at Link\"oping University (Faculty Grant SFOMatLiU No. 2009-00971), from the Knut Alice Wallenberg Foundation (Wallenberg Scholar Grant No. KAW- 2018.0194), as well as support from the Swedish Foundation for Strategic Research (SSF) through the Future Research Leaders 6 program, FFL 15-0290. 
We thank A. Ruban and O. Peil for interesting discussions and for technical help.

\end{acknowledgments}

% \appendix
% \section{Derivation of free energy difference in TTI}\label{app:TTI}

% The free energy of a structure at temperature $T_1$ can be calculated as:
% \begin{equation}
%     \frac{F(T_1,V_2)}{k_B T_1}= \frac{F(T_{\textrm{ref}},V_2)}{k_B T_{\textrm{ref}}}-\int_{T_{\textrm{ref}}}^{T_1} \frac{\langle E \rangle_{T,V_2}}{k_B T^2} \, dT.
% \end{equation}
% If the denominator of the left hand side is brought on the right hand side, and we consider the difference in free energy between the two structures bcc and fcc at the same temperature $T_1$ we obtain:
% \begin{equation}
% \begin{split}
%     &F^{\textrm{fcc}}(T_1,V^{\textrm{fcc}}_2)-F^{\textrm{bcc}}(T_1,V^{\textrm{bcc}}_2)=\\
%     &\left[ F^{\textrm{fcc}}(T_{\textrm{ref}},V^{\textrm{fcc}}_2)-F^{\textrm{bcc}}(T_{\textrm{ref}},V^{\textrm{bcc}}_2) \right] \frac{T_1}{T_{\textrm{ref}}}-\\
%     &T_1\left[ \int_{T_{\textrm{ref}}}^{T_1} \frac{\langle E^{\textrm{fcc}} \rangle_{T,V^{\textrm{fcc}}_2}}{T^2} \, dT - \int_{T_{\textrm{ref}}}^{T_1} \frac{\langle E^{\textrm{bcc}} \rangle_{T,V^{\textrm{bcc}}_2}}{T^2} \, dT \right].
% \end{split}    
% \end{equation}
% From here, derivation of Eq. \ref{eq:TTI} is straightforward.

%\addbibresource{references.bib}
\bibliography{references}

\end{document}